\documentclass[final]{svjour2}
\usepackage{graphicx}
\usepackage{rotating}
\usepackage{amssymb}
\usepackage{mathptmx}
\usepackage{lipsum}
\usepackage{subcaption}

\usepackage[font=small,skip=-10pt]{caption}

\usepackage{siunitx} 
\usepackage{amsmath}

\usepackage[colorlinks=true,urlcolor=blue, citecolor=red]{hyperref}%
\usepackage[colorinlistoftodos]{todonotes}

\usepackage{xspace} 
\newcommand{\sinx}{SiN$_\textrm{x}$\xspace}

\makeatletter
\journalname{Journal of Low Temperature Physics}

\begin{document}

\title{Design and performance of the antenna-coupled lumped-element kinetic inductance detector}

\author{P. S. Barry $^{a,b}$, S. Doyle $^{b}$, A. L. Hornsby $^{b}$, A. Kofman $^{c}$,  E. Mayer $^{a}$, Q. Y. Tang $^{a}$, J. Vieira $^{c}$ and E. Shirokoff $^{a}$}

\institute{$^a$ Kavli Institute for Cosmological Physics, University of Chicago, Chicago, IL\\
\email{ \href{mailto:barryp@uchicago.edu}{barryp@uchicago.edu} } \\
$^{b}$ School of Physics and Astronomy, Cardiff University, The Parade, Cardiff, CF24 3AA, U.K.\\
$^{c}$ Astronomy Department, University of Illinois, 1002 W. Green St., Urbana, IL 61801
}
\date{XX.XX.20XX}

\maketitle

\begin{abstract}
Focal plane arrays consisting of low-noise, polarisation-sensitive detectors have made possible the pioneering advances in the study of the cosmic microwave background (CMB). To make further progress, the next generation of CMB experiments (e.g. CMB-S4) will require a substantial increase in the number of detectors compared to the current stage 3, instruments. Arrays of kinetic inductance detectors (KIDs) provide a possible path to realising such large format arrays owing to their intrinsic multiplexing advantage and relative cryogenic simplicity. In this proceedings, we report on the design of a novel variant of the traditional KID design; the antenna-coupled lumped-element KID. A polarisation-sensitive twin-slot antenna placed behind an optimised hemispherical lens couples power onto a thin-film superconducting microstrip line. The power is then guided into the inductive section of an aluminium KID where it is absorbed and modifies both the resonant frequency and quality factor of the KID. We present the various aspects of the design and preliminary results from the first set of seven-element prototype arrays and compare to the expected modelled performance. 
%
%
%
%
%
%
%
%

\end{abstract}
\section{Introduction}
Kinetic inductance detectors (KIDs) \cite{Day:2003hh} are now considered a compelling alternative technology for large-format focal plane arrays. A number of projects operating at millimetre (NIKA-2 \cite{Calvo:2016ct}, TolTEC \cite{toltec}, SuperSpec \cite{Shirokoff:2014gu,Wheeler:2016dr}) and submillimetre (SpaceKIDs \cite{Griffin:2016kv}, BLAST-TNG \cite{Galitzki:2014ek}) wavelengths will help to deliver the first large scale on-sky demonstration of KID arrays, whose promise of high multiplexing factors and simple fabrication provide a path to reduced cost and complexity of future large-format focal plane instruments.

However, to date, experiments designed to measure the cosmic microwave background (CMB) have been led by instruments based on arrays of transition edge sensors (TESs). State-of-the-art CMB experiments are at 'stage 3', with focal planes containing 5-15k TES detectors \cite{2014SPIE.9153E..1NA,Benson:2014br,Thornton:2016ks}, each approaching the background limit. In this case, further improvements in sensitivity can only be achieved by increasing total focal plane area and number of detectors. Furthermore, motivated by the desire to constrain foreground contamination, current experiments have implemented integrated filtering schemes, where each pixel in the focal plane accepts a wide-band signal that is divided into multiple sub-bands with lithographic on-chip filters and sensed with separate detectors. The required increase in the number of detectors and corresponding readout needed to achieve the goals of the future experiments is a significant technological challenge. Estimates based on background-limited detectors indicate that CMB-S4 will require a minimum of 500k detectors across $\sim10$ ground-based telescopes \cite{Abazajian:2016vh}. It is here where the advantages of KIDs could play an important role in realising the next generation of CMB experiments.  

A KID is a superconducting, high-Q resonator that senses the change in complex surface impedance of the superconducting film upon photon-absorption. Generally, there are two implementations of thin-film superconducting resonators used as the KID transducer; distributed resonators (mKIDs) \cite{Day:2003hh} that are based on half-, or quarter-wave planar transmission line resonators, and lumped-element resonators (leKIDs), that combine discrete and separate inductive and capacitive elements to form a standard LC circuit \cite{Doyle:2008gc}. For filled array architectures, the advantage of the leKID derives from the arrangement of the inductive element of the resonator into an efficient free-space absorber. As the current distribution is constant across the inductor, the need for the external radiation coupling scheme is removed. Furthermore, the ability to separate the inductive and capacitive elements provides an additional degree of versatility that can be used to optimise the device geometry for a specific application.

Despite these advantages, the traditional leKID design is not immediately compatible with the chip-level filtering schemes envisioned for future CMB experiments. In this paper we report on the design of the antenna-coupled leKID, which provides a method to combine the advantages of the traditional leKID design with on-chip band definition to construct highly-multiplexed, polarisation sensitive, multi-chroic focal planes.

\section{mm-Wave Coupling}

\begin{figure}
\begin{center}

\begin{subfigure}[t]{0.3\textwidth}
	\includegraphics[width = \textwidth, keepaspectratio]{cmbkids_singlepol_schem.pdf}
    \label{fig:pixel_schematic}
\end{subfigure}%
\begin{subfigure}[t]{0.35\textwidth}
	\includegraphics[width = \textwidth, keepaspectratio]{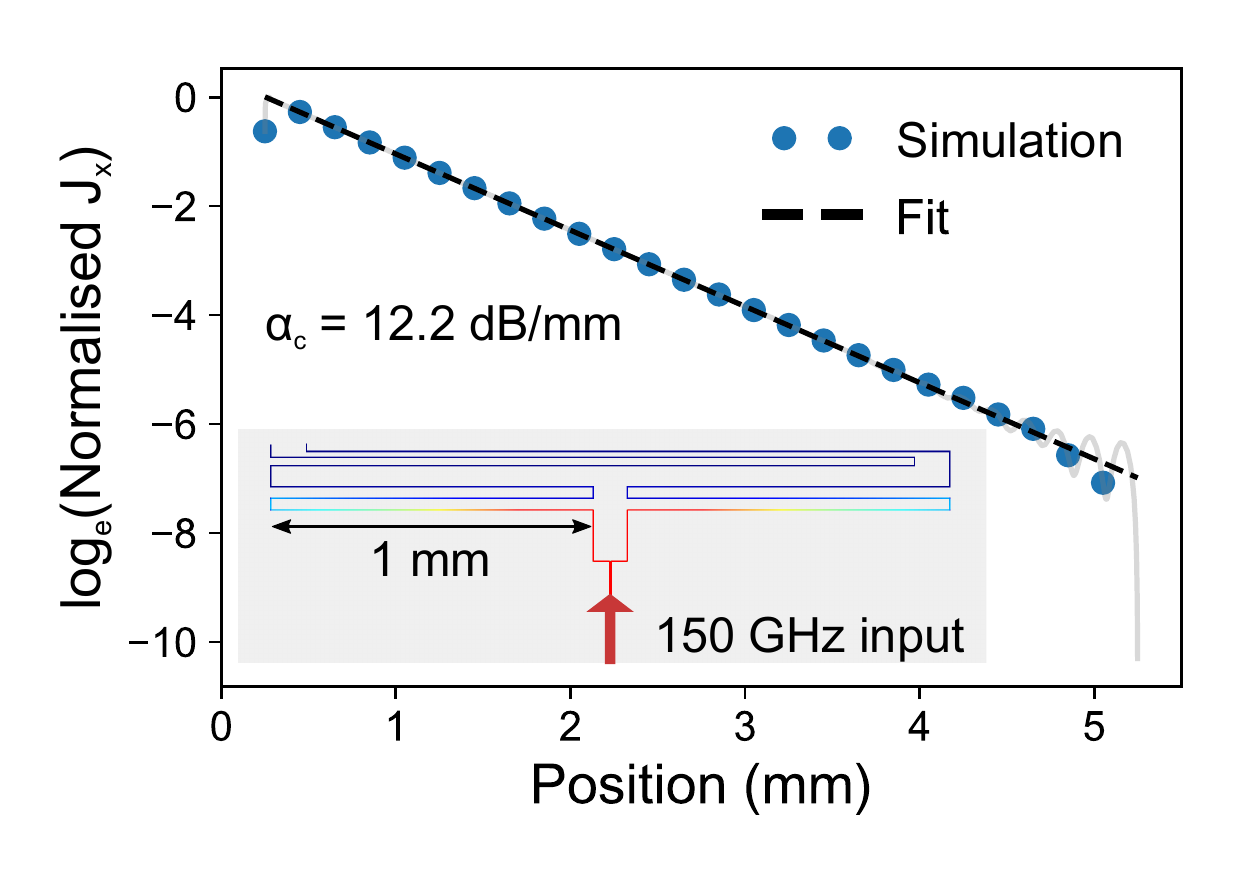}
    \label{fig:jxy_sim}
\end{subfigure}%
\begin{subfigure}[t]{0.35\textwidth}
	\includegraphics[width = \textwidth, keepaspectratio]{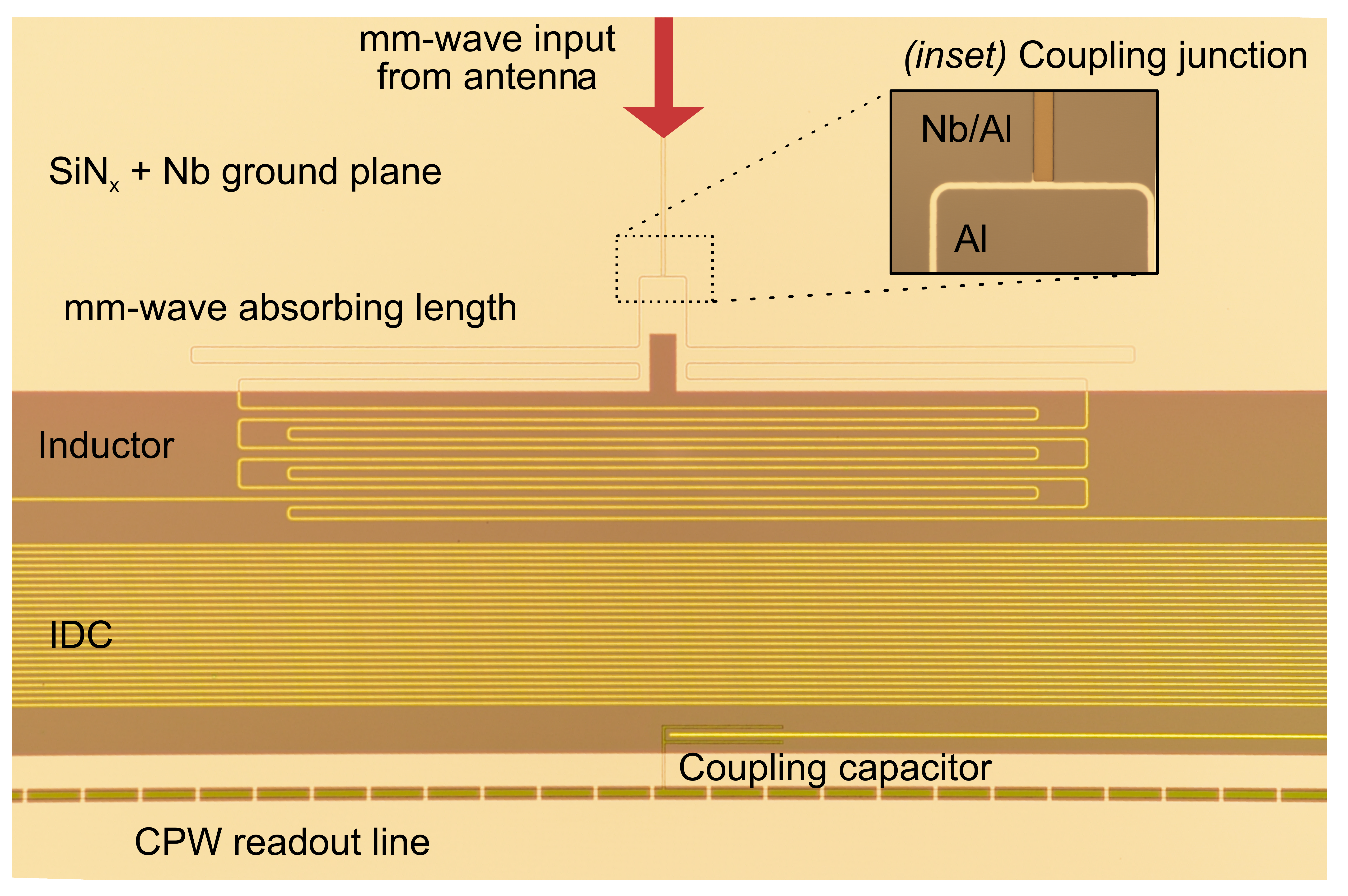}
    \label{fig:pixel_closeup}
\end{subfigure}%
\end{center}
\caption{(Color online) \textit{left}) Schematic of a single pixel from the prototype array, \textit{middle}) Simulated current density along the lossy Al microstrip line at 150 GHz, \textit{right}) Optical microscope image of the first fabricated arrays.}
\label{fig:1}
\end{figure}

At the heart of the design of the antenna-coupled leKID is the coupling mechanism from the mm-wave feedline into the detector. Radiation is coupled onto a thin-film superconducting microstrip line using a lens-coupled twin-slot antenna. We use an inverted  microstrip design \cite{Shirokoff:2014gu}, where the circuit layer is deposited onto a clean, high-resistivity silicon surface, and is covered by the dielectric and ground plane layers, shown in Fig. \ref{fig:1}. The principal advantage in adopting this design is that deposition of the detector Al layer is the first process in the fabrication procedure. This allows maximal control over the substrate preparation which has been shown to be critical to minimizing the contributions to the loss and noise originating from two-level fluctuations. Full details on the fabrication procedure are given in Tang et al. \cite{amyltd2017}. 

The microstrip line is made from an Al/Nb bilayer of thickness \SI{50/250}{\nano\metre}, with $T_c =$ \SI{6.8}{\kelvin} ($\nu_{\textrm{gap}}\approx$ \SI{500}{\giga\hertz}). The bilayer is used to simplify the fabrication of the mm-wave-to-KID junction.  A \SI{500}{\nano\metre} PECVD \sinx layer forms the dielectric insulator and a final \SI{250}{\nano\metre} Nb layer acts as the groundplane, into which the antenna slots are patterned. The microstrip linewidth should remain small in order to reduce radiation losses, and is largely determined by lithographic tolerances. For our design we use a \SI{4}{\micro\metre} microstip line that results in characteristic impedance of $Z_{0} = $\SI{17.6}{\ohm}. The microstrip from the two slots is then combined into a single line that feeds the leKID. In the interest of understanding the fundamental device performance, we have chosen to not include any band-defining components in the prototype devices. However, it is trivial to add any planar filter into the design at this point, and is something we are working to include subsequent device iterations. 

At the leKID, the Al/Nb microstrip line is split and continues as an Al-only microstrip line (cf. inset of Fig. \ref{fig:1}) that doubles as the leKID inductor. The detector is patterned into the base Al layer and provides a simple way to achieve a low-loss galvanic contact between the input transmission line and leKID inductor. From a mm-wave perspective, the leKID inductor acts as a low-loss transmission line with an impedance chosen to ensure the input impedance looking into the inductor is matched (i.e. $Z_{0,\textrm{out}} = 2Z_{0,\textrm{in}}$). As this junction is non-resonant, it is inherently wide-band. However, because the Al microstrip has non-zero loss at mm-wave frequencies above the gap ($\nu_{\textrm{gap}}\approx$ \SI{100}{\giga\hertz}), the characteristic impedance has a non-zero imaginary component, which presents a small impedance mismatch at the junction. 

Planar EM simulations suggest that the match can be better than 20 dB cross the entire millimetre-wave band (90-300 GHz). 
The simulation defines each metal layer by a sheet impedance, which for Al is well characterised at frequencies both below and above the gap frequency. For modelling the Al/Nb bilayer we assume that the sheet impedance is similar to bare Nb. While we expect this assumption to be reasonable, we are in the process of measuring this quantity to inform future design iterations. Furthermore, we expect that the Nb layer will proximitise the Al in the vicinity of the junction and provide a smoother impedance transition relative to our current simulations. 

Once the mm-wave power is travelling in the inductor, the amplitude is expected to decrease exponentially along the inductor length. To ensure that all of the power is absorbed in the inductor, the length of the absorbing section must be chosen appropriately. The attenuation length ($\alpha_c$) for the Al microstrip line can be estimated using the standard expression for microstrip conductor loss in \si{dB\per\milli\metre},
\begin{equation}
\alpha_c = 8.686 \cdot \frac{R_s}{2 Z_0 w}.
\end{equation}
For $R_{s}^{\textrm{Al}} = $ \SI{0.2}{\ohm \per sq} which we derive for a 50 nm film, $Z_0 = \SI{17.6}{\ohm}$, and linewidth $w =$ \SI{4}{\micro\metre}, we estimate $\alpha_c \approx$ \SI{12.3}{dB\per\milli\metre}. Comparison with a 2D simulation of the current density profile gives value of \SI{12.2}{dB\per\milli\metre} (cf. Fig \ref{fig:1}), in good agreement with the analytic estimate. This value of $\alpha_c$ suggests that a length of greater than a couple of millimetres will be sufficient to absorb the incoming signal. For the prototype design, we allow \SI{3}{\milli\metre} length in each arm before the microstrip line is terminated by removing the ground-plane and \sinx. From the mm-wave perspective, the abrupt impedance mismatch will reflect any power that reaches the termination, and provide a second pass for absorption. 

\section{RF Design}
The design of the leKID is subject to a number of constraints unique to this architecture. As discussed above, the portion of the inductor where the mm-wave absorption takes place remains under the ground plane and acts as a lossy microstrip line, absorbing and attenuating the mm-wave signal along the length. The rest of the inductor, and inter-digital capacitor (IDC), is uncovered, as shown in Fig. \ref{fig:1}. The motivation behind removing the groundplane and dielectric above the length of inductor not acting as an absorber, is to reduce the distributed capacitance to ground, which would otherwise dominate the overall capacitance of the resonator and modify the current distribution across the inductor. Furthermore, the dielectric above the IDC is removed to reduce the effect of TLSs loss and fluctuation noise.

The mm-wave input microstrip is connected to the centre of the inductor at the voltage node of the resonator. Placing the microstrip at this location is not expected to affect the RF properties of the resonator as there is negligibly small voltage to drive any appreciable current out of the resonator. Moreover, the direct capacitive connection to ground can potentially enhance the uniformity of resonator frequencies by reducing the effect of variable stray capacitances to external circuit components (e.g. sample box). 

\begin{figure}
\begin{center}
\begin{subfigure}[t]{0.45\textwidth}
	\includegraphics[width = \textwidth, keepaspectratio]{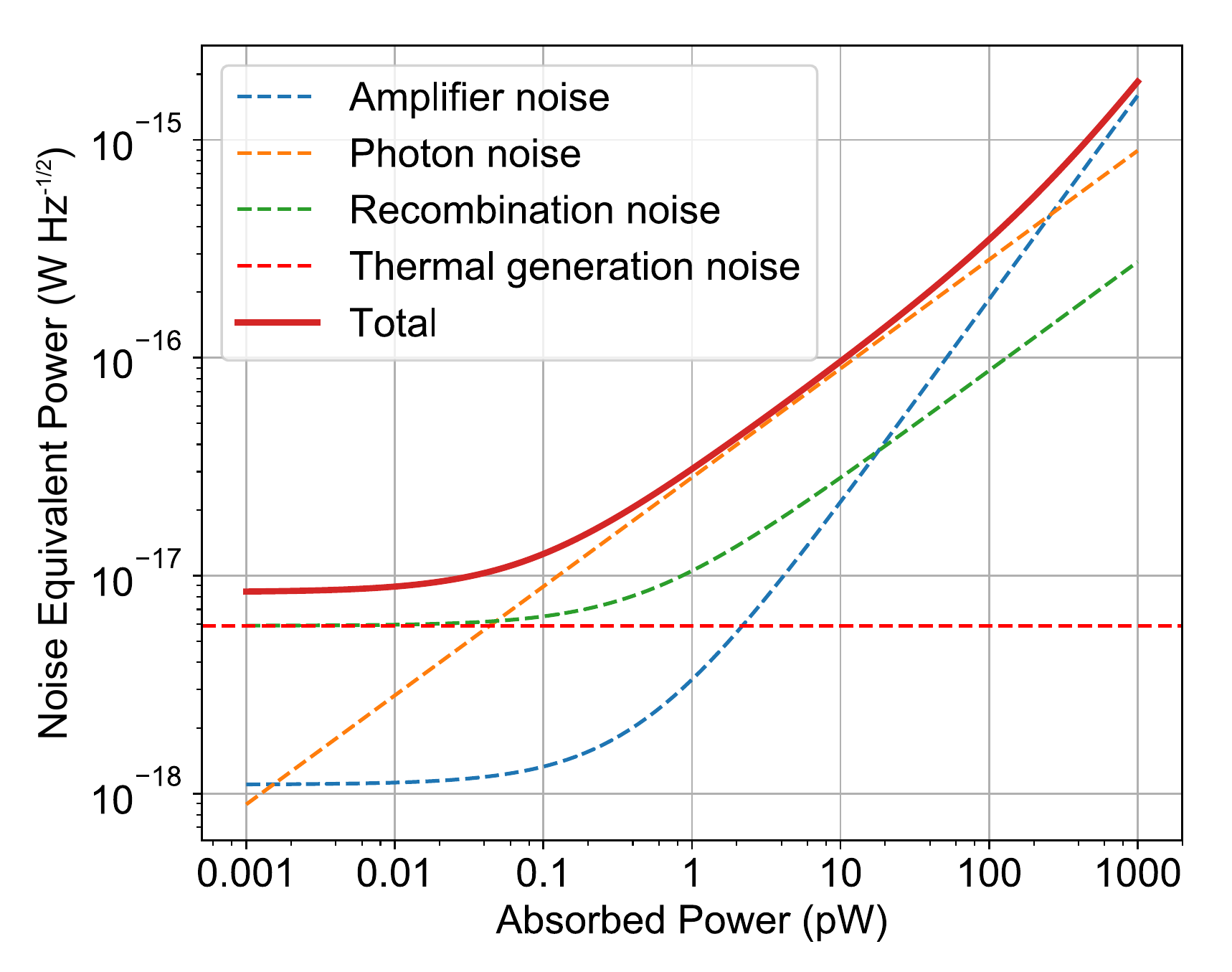}
    \label{fig:nep_model}
\end{subfigure}%
\begin{subfigure}[t]{0.54\textwidth}
	\includegraphics[width = \textwidth, keepaspectratio]{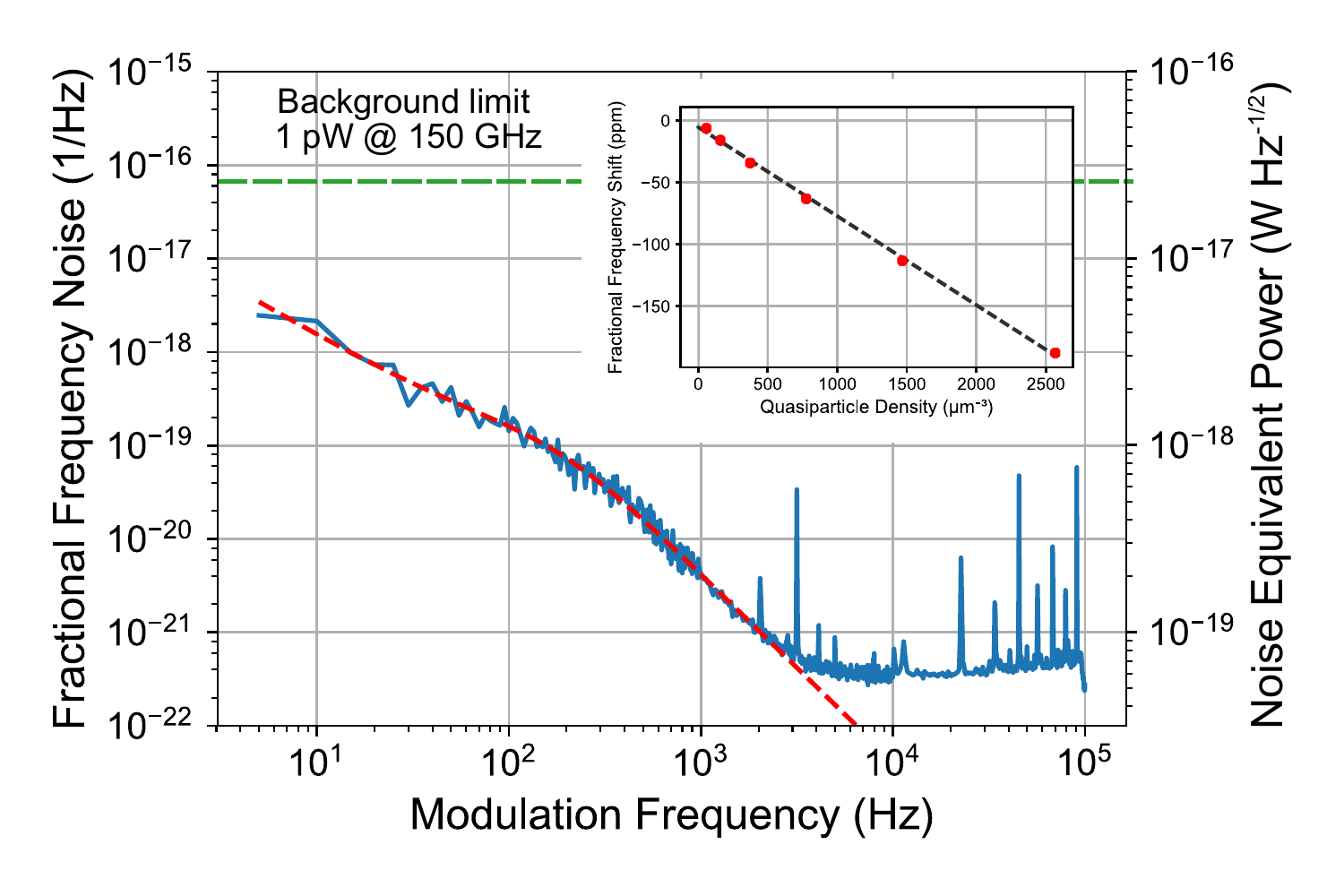}
    \label{fig:nep_meas}
\end{subfigure}%
\end{center}
\caption{(Color online) left) Calculated expected contributions to the noise equivalent power (NEP) as a function of optical load at $T_b =$ \SI{230}{\milli\kelvin}, right) Measured noise spectrum at 100 mK with a fit (red dashed line) to extract the resonator time constant. Right axis is an estimate of the electrical NEP using a measure of the dark responsivity (inset). Dashed green line is the expected background limited NEP for \SI{1}{\pico\watt} optical load at 150 GHz.}
\label{fig:2}
\end{figure}

We calculate the expected sensitivity using a modified responsivity model based on the total number of quasiparticles. We assume a optical loading typical of ground-based CMB experiments of between \SIrange[range-units=single]{1}{10}{\pico\watt} \cite{2015ApJ...812..176B}, which sets the background-limited $\textrm{NEP}_{\gamma} = 2h\nu P_{abs}\left( 1 + \bar{n} \right)$. From the expected power loading, we calculate the total number of quasiparticles that includes, self-consistently, contributions from both the optical signal and thermal excitations present at a base temperature of \SI{250}{\milli\kelvin}. The left panel of Fig. \ref{fig:2} shows the various contributions to the noise for a nominal volume of \SI{4800}{\micro\metre\cubed}, and indicates that the device should remain background-limited for the expected optical loading.  

As only a portion of the inductor is used for mm-wave absorption, the amount of additional inductance will act to reduce the responsivity. The normal method to account for this reduction would be incorporate a partial kinetic inductance fraction ($\alpha^*_k$). However, in this case, the calculation of $\alpha^*_k$ is not straightforward, as quasiparticles generated in the mm-wave absorbing section are able to diffuse to the remainder of the inductor, and still contribute to the responsivity. The diffusion length is expected to be on the order of 1 mm \cite{Hsieh:1968jm} for typically measured quasiparticle lifetimes in Al films. Therefore, we have included a range of inductor volumes around the nominal value that will allow us to quantify the magnitude of this effect by measuring the optical responsivity. 

\section{Preliminary Results}

In this section we present a preliminary characterisation of the microwave performance of the detectors tested in a dark environment. We place the devices into a gold-plated OFHC copper sample box, mounted in-situ with the lens array using our novel alignment scheme, as described in Tang et al. \cite{amyltd2017}. The box is mounted onto the mixing chamber of a dilution refrigerator with a base temperature of 10 mK. The resonant frequency and quality factor are extracted from fits to VNA sweeps, and a standard single-tone homodyne configuration is used to characterise the noise performance \cite{Day:2003hh}. 
 
We measure internal quality factors ($Q_i$) in excess of 500k for all resonators without the \sinx layer over the capacitors \cite{hornsbyltd2017}. This is an important result for this design as it demonstrates that placing \sinx onto the inductive section of a leKID does not have a significant impact on $Q_i$. In terms of scaling to large arrays, $Q_i > 2\times10^5$ is well within the requirements for multiplexing ratios of above $10^3$ using existing readout hardware \cite{Gordon:2016de}. 

To estimate performance of the prototype device from a detector perspective, we measure the electrical noise equivalent power (NEP). It has recently been shown that an estimate of electrical NEP provides an accurate indication of the optical performance for devices where the contribution to the responsivity is constant across the absorbing volume \cite{Janssen:2014eo}. For a leKID, this is automatically satisfied as the current density across the inductor is constant.

The electrical NEP is calculated from Eq. \ref{eq:eNEP}. An estimate of the responsivity is determined from a fit to the resonator response as a function base temperature. The noise level ($S_{xx}$) and quasiparticle lifetime ($\tau_{qp}$) are evaluated from the noise spectrum, and along with known material and geometric parameters, the NEP is calculated from \cite{Baselmans:2008kd}
\begin{equation}\label{eq:eNEP}
\textrm{NEP}_{\delta x} = \sqrt{S_{xx}} \cdot \left[ \frac{dx}{dn_{qp}} \cdot \frac{\eta_{pb}\tau_{qp}}{\Delta_0 V_L} \right]^{-1},
\end{equation} 
where $x= \delta f/f_r$ is the fractional frequency shift, $\eta_{pb}=0.57$ is the pair breaking efficiency, $\Delta_0$ is the low temperature superconducting gap, and the equilibrium quasiparticle density is given by 
\begin{equation}
n_{qp} = 2 N_0 \sqrt{2 \pi k_B T_c \Delta_0 }\exp{(-h\nu/k_B T)}.
\end{equation}
The inset of the right panel of Fig. \ref{fig:2} shows the measured $\delta x$ vs $n_{qp}$ for a representative resonator. The fit range is chosen to capture only the higher temperature behaviour where the quasiparticle response dominates over the dielectric response. A linear fit to the response is used to extract the quasiparticle responsivity. The quasiparticle lifetime is extracted from a fit to roll-off in the noise spectrum. The measured $\tau_{qp} =$ \SI{850}{\micro\second} at 100 mK, which is significantly longer than the expected resonator ring-down time $\tau_{res} = 2Q_r / \omega_0 \approx$ \SI{50}{\micro\second}. 
%
%

The right panel of Fig. \ref{fig:2} shows the measured noise spectrum for a resonator at \SI{100}{\milli\kelvin}. The $1/f^n$ spectral shape indicates that the noise performance at low modulation frequencies is most likely limited TLS fluctuation noise. However, the noise level is well below the expected background limit at \SI{1}{\pico\watt} down to \SI{1}{\hertz}. In subsequent devices we will prioritise further reducing this level by modifying the capacitor geometry and improving the substrate preparation techniques.

The NEP derived from the responsivity is shown on the right-hand y-axis. The dashed line is the expected background limited NEP for the expected detector loading of \SI{1}{\pico\watt}, and indicates that at the expected loading from a ground-based CMB experiment, the performance of these devices will be firmly background limited, with a device-noise $1/f^n$ knee below 1 Hz. 

\section{Conclusions and Future Work}
We have introduced the concept and design of the antenna-coupled leKID, which will enable the advantages of the leKIDs to be combined with existing multi-chroic on-chip mm-wave filtering circuits that will be crucial for future CMB experiments. The measured electrical NEP is promising, with a significant amount of clearance to the expected level for background limited sensitivity. Based on this data, this detector architecture appears to be an interesting candidate for ground based CMB experiments with detector power loading of 1-10 pW. The addition of optical data and characterisation will determine the ultimate performance of this device, and we plan to report on this soon.

\begin{acknowledgements}
This material is based in part upon work supported by the National Science Foundation under Grant Number 1554565. This work also made use of the Pritzker Nanofabrication Facility of the Institute for Molecular Engineering at the University of Chicago, which receives support from SHyNE, a node of the National Science Foundation’s National Nanotechnology Coordinated Infrastructure (NSF NNCI-1542205). We acknowledge support from the Science and Technology Facilities Council (STFC) Consolidated grant Ref: ST/N000706/1 for supporting this work in the UK. 
\end{acknowledgements}

\end{document}